\documentclass[12pt,preprint]{aastex}

\def\gtorder{\mathrel{\raise.3ex\hbox{$>$}\mkern-14mu
                \lower0.6ex\hbox{$\sim$}}}
\def\ltorder{\mathrel{\raise.3ex\hbox{$<$}\mkern-14mu
                \lower0.6ex\hbox{$\sim$}}}

\begin{document}

\title{Ionizing Photon Emission Rates from O- and Early B-Type
Stars and Clusters}
\vspace{1cm}
\author{Amiel Sternberg}
\vspace{0.5cm}
\affil{School of Physics and Astronomy and the Wise Observatory,
        The Beverly and Raymond Sackler Faculty of Exact Sciences,
        Tel Aviv University, Tel Aviv 69978, Israel}
\vspace{0.5cm}
\author{Tadziu L. Hoffmann and A.W.A. Pauldrach}
\vspace{0.5cm}
\affil{Institute for Astronomy and Astrophysics of the Munich
  University, Scheinerstra\ss\/e 1, D-81679 Munich, Germany}
\vspace{1cm}

\begin{abstract}
We present new computations of the ionizing spectral energy distributions
(SEDs), and
Lyman continuum (Lyc) and He {\small I}
continuum photon emission rates, for hot O-type and early
B-type stars. We consider solar metallicity 
stars, with effective temperatures
ranging from 25,000 to 55,000 K, and surface gravities
(cm s$^{-2}$)
log $g$ ranging from 3 to 4, covering the
full range of spectral types and luminosity classes
for hot stars.  We use
our updated ({\it WM-basic}) code to
construct radiation-driven wind atmosphere models
for hot stars. Our models include the
coupled effects of hydrodynamics and non-LTE radiative transfer
in spherically outflowing winds, including
the detailed effects of metal line-blocking and line-blanketing
on the radiative transfer and energy balance.
Our grid of model atmospheres is available on the world-wide-web.
We incorporate our hot-star models into our
population synthesis code (STARS), and we compute the time-dependent
SEDs, and resulting Lyc and He {\small I} emission rates,
for evolving star clusters. We present results
for continuous and impulsive star-formation, for a range of
assumed stellar initial mass functions.

\end{abstract}

\section{Introduction}
\label{introduction}

Massive hot stars inject large amounts of energy and entropy into
the interstellar medium of galaxies, 
via jets, winds, radiation, and supernova explosions.
The production rates of Lyman continuum photons
(Lyc; $h\nu > 13.6$ eV), and of
helium continuum photons (He {\small I}; $h\nu > 24.6$ eV),
are particularly important and 
fundamental parameters of O-type and early B-type stars
(Mihalas 1967; Panagia 1973; Vacca, Garmany \& Shull 1996; 
Schaerer \& de Koter 1997; Lanz \& Hubeny 2003). 
The ionizing radiation produces H{\small II} regions 
around the hot stars, and the photon fluxes determine the intensities
of hydrogen and helium recombination lines, 
collisionally excited metal emission lines, and free-free continuum fluxes
from the photoionized gas.   Massive stars often form in groups or
clusters, as in Galactic OB associations
(McKee \& Williams 1997), or in more
massive super-star-clusters observed 
in galaxy nuclei and starburst galaxies 
(Whitmore et al.~1999; Maoz et al.~2001). Such objects can contain thousands of
individual OB stars within small volumes, and
the associated nebular emissions are detectable to large redshifts
(e.g.~Shapley et al.~2003).

In this paper, we present new computations of the 
Lyc and He {\small I} photon emission rates for 
hot-stars, using our updated radiation-driven wind atmosphere models
for O-type and early B-type stars
(Pauldrach, Hoffmann \& Lennon 2001).
Our models include the
coupled effects of hydrodynamics, and NLTE radiative transfer,
in spherically outflowing radiation-driven winds, including
the detailed effects of line-blocking and line-blanketing
on the radiative transfer and energy balance.
In this paper we
focus on solar metallicity systems, and we present results
for a model grid spanning a wide range of
stellar effective temperatures and surface gravities,
covering the full range of stellar types, from dwarfs to supergiants.
Low metallicity systems such as those occurring in dwarf galaxies
(e.g.~Lee, Grebel \& Hodge 2003) or in the early universe
(Tumlinson \& Shull 2000; Schaerer 2003) will be considered elsewhere.

We use our stellar atmosphere grid,
in combination with evolutionary models,
to carry out ``population synthesis'' computations 
(Bruzual \& Charlot 1993; Leitherer \& Heckman 1995;
Sternberg 1998; Leitherer et al.~1999)
of the time-dependent spectral energy distributions (SEDs), and
time-dependent Lyc and He {\small I} photon emission rates, 
for evolving star-clusters.
We present results for the two limiting cases of 
``continuous'' and ``impulsive'' star-formation,
for a range of assumed stellar initial mass functions.
Our cluster synthesis computations
may be used together with observational continuum and emission-line
data to infer star-formation rates and histories
in galactic and extragalactic star-forming regions.

In \S 2, we  summarize the basic ingredients of our
stellar atmosphere models. In \S 3, we fix the various
empirically based input parameters, including the
stellar radii, surface gravities,
effective temperatures, terminal wind
velocities, and mass-loss rates. In \S 4, we discuss
the model atmospheres, present 
the Lyc and He {\small I}
photon emission rates as functions of the stellar parameters, 
and compare our results to previous
computations presented in the literature.  In \S 5,
we present the results of our population synthesis
computations for cluster SEDs and associated
Lyc and He {\small I}
photon emission rates.

\section{Model Atmospheres}

We use our  NLTE wind-driven
stellar atmosphere code ({\it WM-basic})
to compute hot-star spectral energy distributions 
for a wide range of stellar input parameters. 
Details of this code, and the modelling methods, are presented in
Pauldrach et al.~(2001). Here we provide a brief description.

The {\it WM-basic} code iteratively solves the coupled equations
of hydrodynamics, energy balance, and NLTE radiative transfer, 
for spherically symmetric radiatively driven steady-state winds 
expanding off the photospheres of hot stars.  
The hydrodynamic equations are solved for pre-specified
values of the stellar effective temperature
$T_{\rm eff}$, surface gravity log $g$, photospheric 
radius $R_*$, and metallicity $Z$.
Given the solutions to the hydrodynamic density and velocity structures, the 
local NLTE occupation numbers for approximately 5000 atomic levels are 
determined by solving the microscopic population rate equations, including
radiative and collisional excitation and deexcitation processes, in
several million line-transitions of elements from H to Zn. 
The strong ``line blocking'' attenuation of the ultraviolet (UV) radiation
fluxes (both ionizing and non-ionizing) by the numerous
spectral lines in the UV spectral range  is properly
included in the radiative transfer, with significant
influence on the NLTE level populations.  The ``line blanketing'' heating of
the photosphere as a consequence of this
blocking of radiation
is accounted for in the
temperature structure, which is determined by solving the energy balance
equation.  The radiation field, opacities, emissivities, and
occupation numbers at all depths in the wind are determined
consistently via a detailed computation of the radiation
transfer in the observer's frame, and an explicit solution of
the rate equations incorporating all significant microscopic
processes.  The solution to the radiative transfer equation
finally also yields the escaping radiation field in the
form of the SED and
synthetic stellar spectrum.               

The steady-state flow, in both the supersonic and subsonic
regions, is controlled by the
radiative acceleration provided by the combined
effects of continuum
and line absorption. In the computations we present here,
we adjust the absorption line ``force multiplier'' parameters,
$k$, $\alpha$, and $\delta$ (Castor, Abbott \& Klein 1975;
Abbott 1982), such that
the resulting hydrodynamics mass-loss rates, $\dot{M}$,
and terminal flow velocities, $v_\infty$,
are consistent with observations (see \S 3).

Our ``realistic'' NLTE radiation-driven wind atmosphere models
provide excellent matches to directly observable far-UV 
stellar SEDs longward of the Lyman limit
(e.g., Pauldrach et al.~2001). Crucially, the computed far-UV spectra
depend on the detailed NLTE solutions throughout the winds,
which, in turn, depend primarily on the ionization states
of species with ionization potentials {\it shortward} of
the Lyman limit. Thus, the excellent match to observed far-UV
SEDs is direct evidence that the ionizing
continua are computed accurately in our models,
and that additional physical processes, such as gas clumping,
have only minor effects. Further 
supporting evidence for the accuracy of our models comes from successful
fits to the relative intensities of (infrared) emission lines
observed in Galactic HII regions, in nebular photoionization calculations
that explicitly incorporate our hot-star models (Giveon et al. 2002).

\section{Stellar Parameters}

We have constructed a detailed grid of OB star model atmospheres
for all luminosity classes, from main-sequence
dwarfs to supergiants.
We adopt solar metals abundances as compiled by Grevesse \& Sauval (1998).
The input parameters for each model are
the effective temperature, 
surface gravity, and stellar radius
{\footnote{Defined as the radius at which the Rosseland
optical depth equals 2/3.}}, as well as the 
terminal wind velocity, $v_\infty$, and mass-loss rate, $\dot{M}$.

We choose parameter values based on observations.
Herrero et al.~(1992) and Puls et al.~(1996) have 
determined the surface gravities, effective temperatures,
and stellar radii, for samples of OB stars, via the
``spectral fitting'' method.
In these studies, log $g$ and $T_{\rm eff}$
were inferred for individual stars
by matching spherical NLTE (but non-blanketed)
atmosphere models to the observed hydrogen 
and helium absorption line profiles in the stellar spectra.
Given the best fitting SEDs and predicted surface radiation flux densities, 
the radii were then inferred given the observed
stellar (V-band) luminosities, following the technique
introduced by Kudritzki (1980).

In Figure 1, we display log $g$ vs.~$T_{\rm eff}$,
and $R_*$ vs.~$T_{\rm eff}$, for the samples of
OB stars analyzed by Herrero et al.~(1992) and Puls et al.~(1996). 
It is evident that the OB stars of various types occupy well defined
regions in these diagrams. The empirical data allow us to define
relations for ``generic'' dwarfs and supergiants. For dwarfs we set
\begin{equation}
{\rm log}\ g = 4
\end{equation}
\begin{equation}
{R\over R_\odot} = 0.2\times T_{\rm eff} + 2 \ \ \ ,
\end{equation} 
and for supergiants,
\begin{equation}
{\rm log}\ g = 0.045 \times T_{\rm eff} + 1.65
\end{equation}
\begin{equation}
{R\over R_\odot} = -0.6 \times T_{\rm eff} + 45 \ \ \ .
\end{equation}
These relations define the approximate boundaries,
in these diagrams, within which
OB stars of all luminosity classes are known to exist.

Our model atmosphere grid covers the full
range of parameters within the ``triangular wedge''
that is defined by the dwarf and supergiant sequences
in the log $g$ vs.~$T_{\rm eff}$ diagram (see Fig. 1a). 
We computed atmospheres for $T_{\rm eff}$ ranging from
25 to 55 kK, in steps of 1 kK, and surface gravities (cm s$^{-2}$)
ranging from (as low as) 3 to 4, in steps of 0.2 dex.
For each $T_{\rm eff}$ and log $g$ pair, we
set the stellar radius via linear interpolation, using log $g$ as
the independent parameter, and  
the dwarf and supergiant radii as specified by equations (1)-(4).
The surface gravities, stellar radii, and effective temperatures,
of our models are displayed in Figures 2a and 2b.
The associated {\it spectroscopic} masses
$M_*\equiv gR^2_*/G$, and bolometric luminosities 
$L_*\equiv 4\pi R^2_*\sigma T_{\rm eff}^4$, for our models are shown in
Figures 2c and 2d.

The terminal wind velocities, $v_\infty$, of OB stars
are directly observable quantities 
(e.g. Prinja, Barlow \& Howarth 1990; Kudritzki \& Puls 2000).
Given the terminal velocities, the mass-loss rates, $\dot{M}$, 
may be determined via the
observed wind-momentum-luminosity (WML) relation between  
the stellar luminosity, and the ``modified wind momentum''
$D_{\rm mom}\equiv \dot{M}v_\infty(R_*/R_\odot)^{0.5}$.
The theory of radiatively driven winds predicts that 
$D_{\rm mom}$ is a simple function of $L_*$.
The empirical WML relation depends on both luminosity
class and spectral type, with scatter.
We use the expressions assembled and summarized
by Kudritzki \& Puls (2000).
Figure 2e shows the resulting empirically based values of
$v_\infty$ vs.~$T_{\rm eff}$ for our low to high
gravity systems.
We  adopt an O-star dwarf/giant WML relation
for stars with $R_* < 15 R_\odot$, and a supergiant WML
relation for stars for $R_* > 15 R_\odot$
\footnote{Following Kudritzki \& Puls we write
${\rm log}D_{\rm mom} = {\rm log}D_0 + x{\rm log}(L/L_\sun)$,
and set ${\rm log}D_0=19.87$ and $x=1.57$ for $R < 15 R_\odot$,
and ${\rm log}D_0=20.69$ and $x=1.51$ for $R > 15 R_\odot$.}.
The resulting mass-loss rates are shown in Figure 2f.
We have verified that the non-physical discontinuity in $\dot{M}$ in the
log $g$=3.8 sequence (see Fig.~2f) that is introduced by our WML {\it Ansatz}
does not lead to significant discontinuities in the properties of
the model atmospheres.

\section{Individual Stars}

In Figure 3 we display a subset of our computed 
stellar atmosphere spectra as functions
of $T_{\rm eff}$ and log $g$, spanning the dwarf to supergiant parameter
space. Our full grid of models is available on the world-wide-web
\footnote{
ftp://wise3.tau.ac.il/pub/stars
}. 
For each model we plot the surface flux density, $f_\nu$,
(erg s$^{-1}$ cm$^{-2}$ Hz$^{-1}$) vs.~the photon energy $E$, from
0 to 5 ryd.  For comparison, for each model 
we also plot the blackbody flux densities, 
$\pi B_\nu(T_{\rm eff})$, where $B_\nu$ is the Planck function. 

The stellar SEDs consist of continua with numerous superposed
absorption and emission features.
Several noteworthy 
properties are apparent in the model atmosphere flux distributions.
First, the enhanced ionizing continua that are
expected in extended and expanding atmospheres 
(Gabler et al.~1989) are moderated and reduced by line-blocking effects 
(Pauldrach et al.~2001). Second,
the prominent Lyman and He {\small I} ionization edges
at 1 and 1.8 ryd disappear as the effective temperature
increases, or as the surface gravity
is reduced. Third, at high $T_{\rm eff}$ and low log $g$,
the SED continua near the ionization edges approach the blackbody curves. 

This behavior may be understood as follows.
Because the continuum on each side of an ionization edge is formed at
a depth and temperature where the
radiation becomes optically thick, and
because of the generally larger opacity on the ``blue'' sides,
prominent edges appear
in systems with steep temperature gradients.
The temperature gradients are, in turn, determined by the associated
hydrodynamical density gradients.  Thus, when radiation pressure
is most effective, as in high $T_{\rm eff}$ or low log $g$
systems, the atmospheres are extended,
the density gradients are diminished, and the ionization edges disappear. 
When this happens,
the continuum at all wavelengths becomes optically thick 
at about the same spatial locations and temperatures.
The overall continuum must then approach the blackbody form,
given the definition of $T_{\rm eff}$ and flux conservation.

Figure 3 also shows that a He {\small II} continuum ($E > 4$ ryd)
is absent in low $T_{\rm eff}$ and low log~$g$ systems (optically thick winds)
but appears at high $T_{\rm eff}$ and high log~$g$ (optically thin winds).
This illustrates the well-known fact that in stellar winds the He~{\small II}
continuum makes a sudden transtion from optically thick to thin behavior
at critical
values of the mean wind density, depending on $T_{\rm eff}$ and the
electron temperature at the base of the wind
(Pauldrach et al.~1990).
The transition from thick to thin winds is also triggered
by an increase in the surface gravity.
The transitions are sudden
because near the critical points an
increased He~{\small II} opacity leads to diminished
metals ionization and increased line blocking in the He~{\small II}
continuum, leading to a
``run-away'' increase in the He~{\small II} opacity
\footnote{
The transition to optically thick winds can also give
rise to enhanced radiative acceleration and a corresponding jump in the mass-loss
rate, possibly leading to bi-stable wind behavior (Pauldrach \& Puls 1990).}.  

For each model in our grid we have computed
the Lyc and He {\small I} flux integrals (s$^{-1}$ cm$^{-2}$)
\begin{equation}
q_{\rm x} = \int_{{\nu_{\rm x}}}^\infty {f_\nu \over h\nu} d\nu
\end{equation}
where $\nu_{\rm x}$ is the ionization frequency for H or He.
In Figures 4a and 4b we plot $q_{\rm H}$ and $q_{\rm He}$ as
functions of $T_{\rm eff}$ for the
different log $g$ sequences, and for comparison we also
plot the blackbody fluxes. It is apparent that
at low $T_{\rm eff}$, both $q_{\rm H}$ and $q_{\rm He}$
are supressed below the blackbody values.
For fixed $T_{\rm eff}$, the fluxes increase and
approach the blackbody curves as log $g$ decreases and
the ionization edges weaken and disappear.
At high $T_{\rm eff}$ the Lyc and He~{\small I} fluxes
approach the blackbody limits.

In Figures 4c and 4d we plot the
Lyc and He {\small I} photon emission rates (photons s$^{-1}$), 
$Q_{\rm H}\equiv 4\pi R_*^2q_{\rm H}$ and
$Q_{\rm He}\equiv 4\pi R_*^2q_{\rm He}$,
as functions of $T_{\rm eff}$ for the various log $g$
sequences, where the stellar radii $R_*$ are as specified in \S 2.
In Figures 4c and 4d we indicate
the locations of dwarf and supergiant stars
as defined by equations (1)-(4). 
The photon emission rates are larger for the
supergiants because of their larger surface areas
and smaller surface gravities.
For $T_{\rm eff}$ from 25 to 50 kK, $Q_{\rm H}$ ranges
from $1\times 10^{46}$ to $5\times 10^{49}$ photons s$^{-1}$
for dwarfs, and from $4\times 10^{47}$ to $8\times 10^{49}$
photons s$^{-1}$ for supergiants. For this temperature range,
$Q_{\rm He}$ ranges from $2\times 10^{43}$ to $1\times 10^{49}$
photons s$^{-1}$ for dwarfs, and from
$1\times 10^{45}$ to $2\times 10^{49}$ photons s$^{-1}$
for supergiants.

Earlier calculations of the Lyc and He {\small I} emission rates
for hot stars are in overall good agreement with our
more sophisticated models, but some important differences
do exist which we now discuss. Specifically,
we compare our Lyc and He {\small I} flux computations to those presented by 
Vacca, Garmany \& Shull (1996), Schaerer \& de Koter (1997),
and Lanz \& Hubeny (2003). 
Vacca et al.~based their computations on static plane-parallel
LTE models (Kurucz 1992). Schaerer \& de Koter carried out computations
using their spherical wind ``{\it CoStar} models'',
in which only hydrogen and helium (but not the metals) are treated in non-LTE,
and in which the temperature structures
are derived using an approximate
treatment of line-blanketing in grey atmospheres.
Lanz \& Hubeny (2003) used
models that do incorporate metals in NLTE, but which are
static and plane-parallel.

Lanz \& Hubeny (2003) provide computations of $q_{\rm H}$ and
$q_{\rm He}$ for a range
metallicities. For solar metallicity stars, their results are
in good agreement with ours at high effective temperatures
where the fluxes approach the black-body curves. However,
for $T\lesssim 40$ kK we find lower Lyc and He~{\small I} fluxes,
with the differences increasing to $\sim 0.5$ dex, in both $q_{\rm H}$
and $q_{\rm He}$, at temperatures below 30 kK. 
 
For purposes of comparison we have
constructed models for the identical stellar
parameters, $T_{\rm eff}$, log $g$
and $R_*$, that were selected by Vacca et al., and also adopted
by Schaerer \& de Koter, 
for stellar luminosity classes V, III, and I.
In Tables 1, 2, and 3, we list these stellar
input parameters, together with
appropriate values for $v_{\infty}$ and
$\dot{M}$, following the procedures described in \S 2.
We also list the spectral-type vs.~effective temperature
calibrations derived by Vacca et al
\footnote{
More recent calibrations for dwarf stars
based on non-LTE wind atmospheres 
(Martins, Schaerer \& Hillier 2002) lower the
temperatures for a given spectral class by a few 10$^3$ K.}.

In Tables 1-3 we list our
results for $q_{\rm H}$, $q_{\rm He}$,
$Q_{\rm H}$ and $Q_{\rm He}$, for the specified stellar parameters.
We compare our results to the Vacca et al.~and Schaerer \& de Koter 
computations in Figure 5. The various computations for the Lyc emission rates
agree very well at high $T_{\rm eff}$ where the
fluxes approach blackbody values. Differences of up to 0.1 dex
are present at low $T_{\rm eff}$, with our values falling
midway between the higher Vacca et al.~and lower Schaerer \& de Koter results
for the giants (III) and supergiants (I). Larger differences
are present in the He {\small I} emission rates. 
For low $T_{\rm eff}$
our results deviate for the Vacca et al.~values by up to
$\pm 0.3$ dex. 
At high $T_{\rm eff}$
our results are about 0.3 dex smaller than Schaerer \& de Koter,
for all luminosity classes. We attribute this difference
to our non-restrictive treatment of NLTE metal line blocking and
blanketing in the hydrodynamics, energy balance, and 
radiative transfer.

\section{Clusters}

We have incorporated our model atmospheres into our
population synthesis code {\rm STARS} (Sternberg 1998;
Thornley et al.~2000) to compute
the time-dependent ionizing spectral energy distributions,
and Lyc and He {\small I} photon emission rates, for
evolving stellar clusters.
Our cluster synthesis computations
may be used together with observations of
stellar continua, and nebular
hydrogen and helium recombination line
data, to infer star-formation rates and histories
in galactic and extragalactic star-forming regions.
 
We use the Geneva stellar evolutionary tracks
for non-rotating stars with solar metallicity and enhanced
mass-loss rates
(Schaller et al.~1992; Meynet et al.~1994; Maeder \& Meynet 2000)
to follow the stellar population distributions in the
Herzprung-Russell (H-R) diagram.
The cluster luminosity density (erg s$^{-1}$ Hz$^{-1}$)
may be written as
\begin{equation}
\label{e:lclus}
L_\nu(t) = \int_{{m_l}}^{{m_u}} \int_{\rm track}
        \rho_{\rm T,L,g} \ \frac{L}{\sigma T_{\rm eff}^4} \
        f_\nu(T_{\rm eff},g) \ ds \ dm \ \ \ ,
\end{equation}
where $\rho_{\rm T,L,g}$ is the number density of stars
per unit ``length''  along evolutionary tracks, 
as specified by the
zero-age main-sequence masses, 
at positions $T_{\rm eff}$ and $L_*$
in the H-R diagram, corresponding to 
{\it evolutionary} surface gravities $g=GM_*/R_*^2$,
where $M_*$ is the (varying) stellar mass along the tracks  
(with $L_*=4\pi R_*^2\sigma T_{\rm eff}^4$). In equation (\ref{e:lclus}),
$f_\nu(T_{\rm eff},g)$ is the SED for individual stars
with temperatures $T_{\rm eff}$ and surface gravities $g$.
The integrations are carried out along the entire tracks for stellar
masses ranging from a lower-limit $m_l$ to an upper-limit $m_u$.

In our computations we assume that stars form with a
Salpeter initial mass-function (IMF), so that the number
of stars formed per mass-interval per unit time
is $n_m(t)=k(t)m^{-\alpha}$, with $\alpha=2.35$,
where $t$ is the age of the stellar cluster.
The total star-formation rate ($M_\odot$ yr$^{-1}$) is then
\begin{equation}
R(t) = k(t) \int_{{m_l}}^{{m_u}}m^{1-\alpha}\ dm \ \ \ .
\end{equation}
We note that 
$\rho_{\rm {T_{\rm eff}},L,g}ds = n_m(\tau)dt$, where the retarded time 
$\tau\equiv t-t_*$, is the lag between the cluster age and
the evolutionary age $t_*$ of a star with temperature $T_{\rm eff}$
and luminosity $L_*$, on a given track.
The ``speed'', $v\equiv ds/dt$, is the
rate at which stars move along the tracks in the H-R diagram.

Our focus here is the computation of the ionizing SEDs,
and resulting total Lyc and He {\small I} photon emission rates, for evolving clusters.
OB stars are the 
primary sources of such radiation, and
for these we use our model atmospheres.
However, we have
also included the contributions of cooler stars (down to 1 $M_\odot$)
in our synthesis computations.
Such stars make a negligible contribution to the
ionizing fluxes, but do eventually dominate the longer
wavelength optical and infrared radiation.
For cool stars we use the Kurucz (1992) 
library of plane parallel LTE atmospheres. 

We adopt a simplified treatment for the likely minor
(but uncertain) contributions of short-lived 
hot Wolf-Rayet (W-R) stars 
(e.g.~Schaerer \& Vacca 1998) to the cluster ionizing SEDs
and total Lyc and He {\small I} photon emission rates.  Recent atmosphere models
(Hillier \& Miller 1998; see also Smith et al.~2002)
for W-R stars, which incorporate the important effects of 
metal line-blocking, suggest that 
He {\small II} continuum photons ($E > 4$ ryd)
are fully absorbed in the dense W-R winds, in both the WC and
WN phases. Between 1 and 4 ryd, the continua of the
model W-R atmospheres are approximately Planckian with 
superposed emission and absorption features, and
with increasing absorption near
the He {\small II} edge (Hillier \& Miller 1998).
The expectation that W-R stars are minor
contributors to the ionizing photon budget in clusters
is supported by observations which show that the presence of such
stars do not influence the associated nebular ionization states
(Bresolin \& Kennicutt 2002; Gonzalez-Delgado et al.~2002). 
Given the uncertainties in the theoretical W-R spectra 
we treat the W-R stars as simple blackbodies between 
1 and 4 ryd, and assume that they
emit no photons beyond the He{\small II} limit.

In our numerical integration of equation (\ref{e:lclus})
we first interpolated a dense set of 641 tracks from the
original set provided by Schaller et al.~(1993).
The interpolated values of log~$L_*$ and log $T_{\rm eff}$
were evaluated between equivalent
evolutionary stages along adjacent tracks.
For given $L_*$ and $T_{\rm eff}$ along a track, we used the
associated (evolutionary) surface gravity
to select the nearest appropriate atmosphere from
our model grid.
The stellar flux densities were then rescaled
by the factor $L_*/\sigma T_{\rm eff}^4$ (see eqn.~[\ref{e:lclus}])
to ensure consistency with the evolutionary luminosities.
In our calculations a ``mass discrepancy''
(e.g.~Groenewegen et al.~[1989]; Herrero et al.~[1992]) 
of up to a factor of $\sim 2$ exists between the
spectroscopic masses as defined in \S 3 (see also
Fig.~[2c]), and the masses as specified by the
evolutionary computations. Corresponding
differences of up to 0.2 dex are thus present between
the ``spectroscopic'' and ``evolutionary'' surface gravities
for given values of $L_*$ and $T_{\rm eff}$. We do
not attempt to resolve this small inconsistency.

In Figure 6 we plot the evolving cluster luminosity densities, for 
continuous and impulsive star-formation, assuming 
Salpeter IMFs with a lower (evolutionary) mass limit of 
$m_l=1$ $M_\odot$, and
upper-mass limits $m_u$ equal to 30 and 120 $M_\odot$.
We display the behavior of the
SEDs from 0.8 to 2.5 ryd. Higher energy photons contribute
negligibly to the total Lyc and He {\small I} emission rates.

For the continuous models we assume a constant
star-formation rate $R=1$ $M_\odot$ yr$^{-1}$
between $m_l$ and $m_u$.
For these models, the luminosity densities initially
increase with cluster age as the numbers of hot stars grow.
The ionizing
continua approach equilibrium at $\sim 10$ Myr
as the total numbers of OB stars in the clusters reach equilibrium.
The SEDs steepen slightly, and the magnitudes of the
Lyman edges correspondingly increase, as the relative numbers
of late vs.~early type O stars increase and
approach equilibrium.

For the impulsive models, we assume that a cluster with total mass
$M_{\rm clus}=10^5$ M$_\odot$ is formed
instantaneously. We again assume a Salpeter IMF
with $m_l=1$ $M_\odot$, and $m_u$ equal to 30 or 120 $M_\odot$. 
For the impulsive bursts
the cluster luminosity densities decay with time as the
most massive and luminous stars disappear from
the systems. The SEDs steepen sharply, and the
Lyman edges grow, as the clusters age and the
hottest stars with the shortest lifetimes disappear.

In Figure 7 we plot the integrated Lyc and He {\small I} photon emission
rates, $Q_{\rm H}$ and $Q_{\rm He}$, for the evolving clusters.
We display results for upper-mass IMF limits $m_u$ equal to
20, 25, 30, 40, 60 and 120 $M_\odot$.  For continuous star-formation,
the emission rates initially
increase linearly with cluster age. Equilibrium is reached
after a time corresponding
to the lifetimes
\footnote{Rotation may increase the stellar lifetimes by
up to $\sim 30\%$ (Meynet \& Maeder 2000).}
of the most massive (and dominating) stars in the
system.  The equilibrium rates are sensitive to the
upper mass limit.
For example,
the asymptotic Lyc emission rates increase by a
factor of 25, from $1.0\times 10^{52}$ to $2.5\times 10^{53}$ s$^{-1}$
(for $m_l=1$ $M_\odot$ and $R=1$ $M_\odot$ yr$^{-1}$)
as $m_u$ is increased from 20 to 120 $M_\odot$.

For impulsive bursts, the photon emission rates are
approximately constant at early times, increase slightly
during the supergiant phases, and then decay rapidly as
the hot stars disappear. The maximal emission rates
are sensitive to $m_u$. The peak Lyc emission rates
increase by a factor of 80, from $1.2\times 10^{50}$ to
$1.0\times 10^{52}$ s$^{-1}$ (for $m_l=1$ $M_\odot$ and
$M_{\rm clus}=10^5$ $M_\odot$) as
$m_u$ is increased from 20 to 120 $M_\odot$.

\section*{Acknowledgments}

We thank I. Hubeny, and T. Lanz for helpful comments, and U. Giveon for
his assistance with the WM-{\it basic} computations.
We thank the German-Israeli Foundation for Research and Development
(grant I-0551-186.07/97) for supporting our research.

\listoffigures

\begin{figure}[p]
\plotone{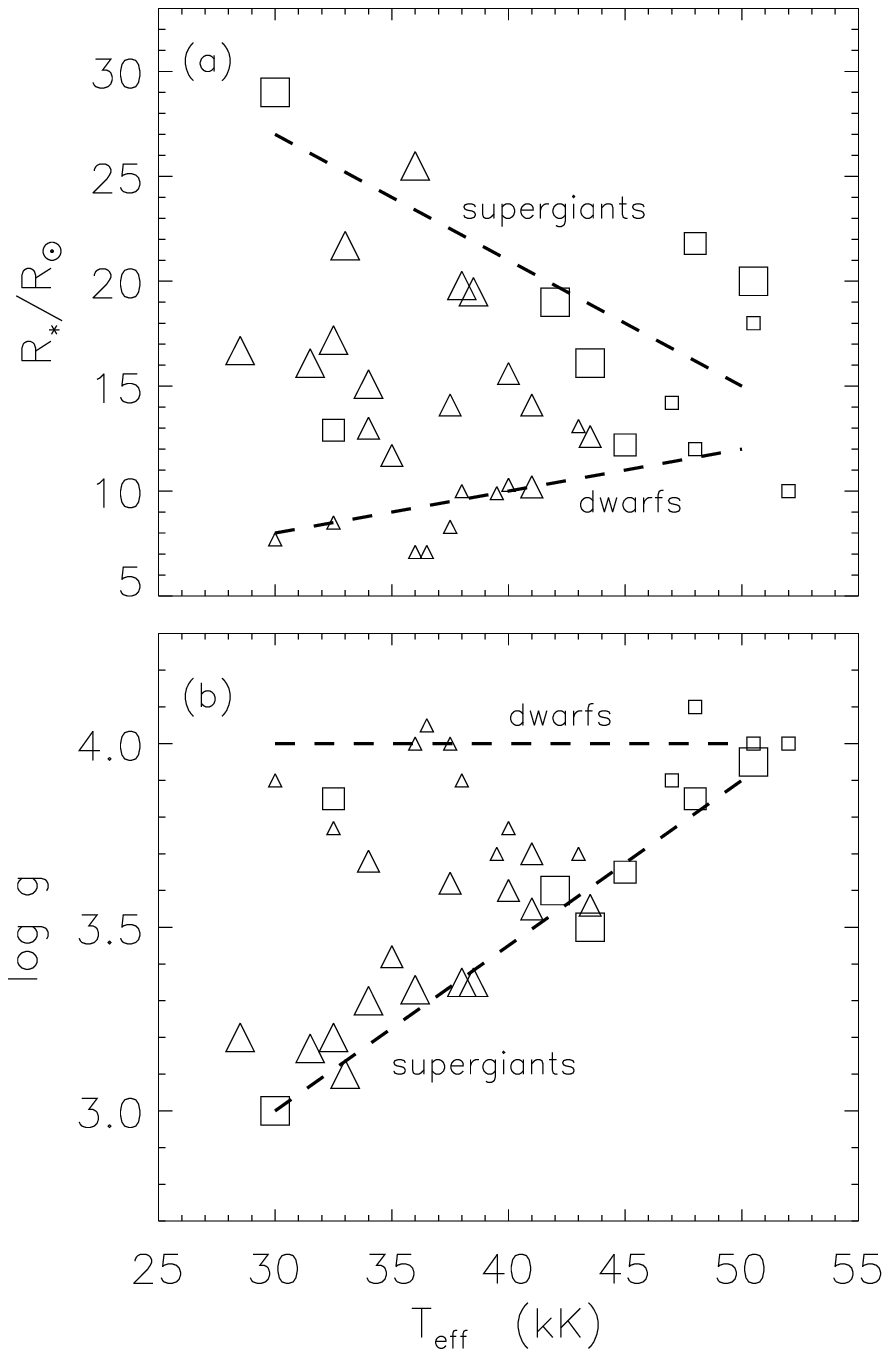}
\vspace{25cm}
\caption[Empirical stellar data, (a)
 stellar radii, and (b) surface gravities,
based on data compiled by Herrero et al.~(1992) (triangles),
and Puls et al.~(1996) (squares). Small, medium, and large
symbols represent dwarf, giant, and supergiant stars.] 
{}
\label{sdata}
\end{figure}

\begin{figure}[p]
\plotone{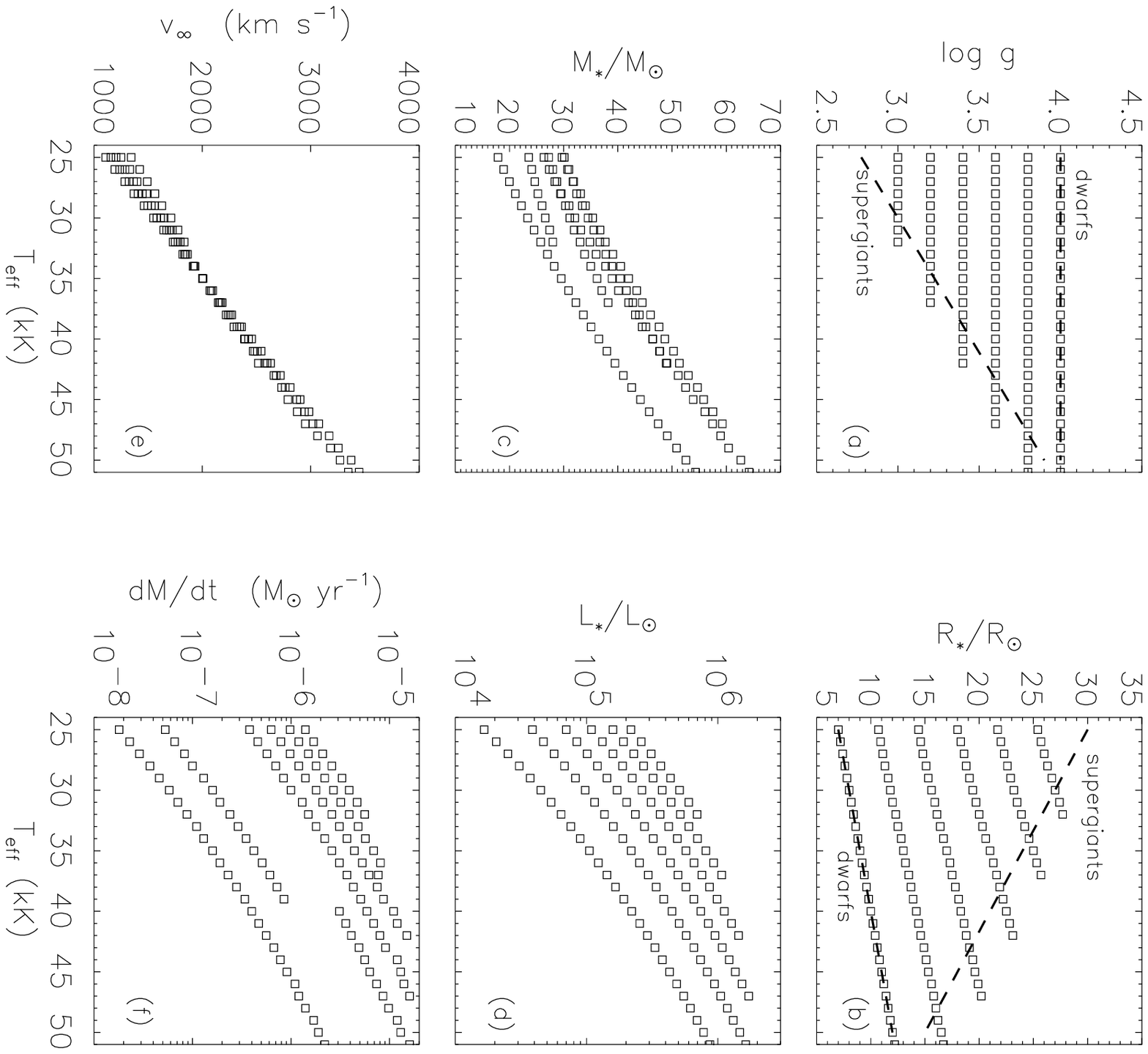}
\caption[Input stellar parameters for the model grid.
(a) surface gravity log $g$, (b) radius $R/R_\odot$, 
(c) spectroscopic mass $M_*/M_\odot$,
(d) luminosity $L_*/L_\odot$, 
(e) terminal wind velocity $v_\odot$ (km s$^{-1}$),
and (f) wind mass-loss rate ${\dot M}$ ($M_\odot$ yr$^{-1}$).
The dashed lines in panels (a) and (b) are the dwarf and
supergiant sequences as defined by equations (1)-(4).] 
{}
\label{inputs}
\end{figure}

\begin{figure}[p]
\plotone{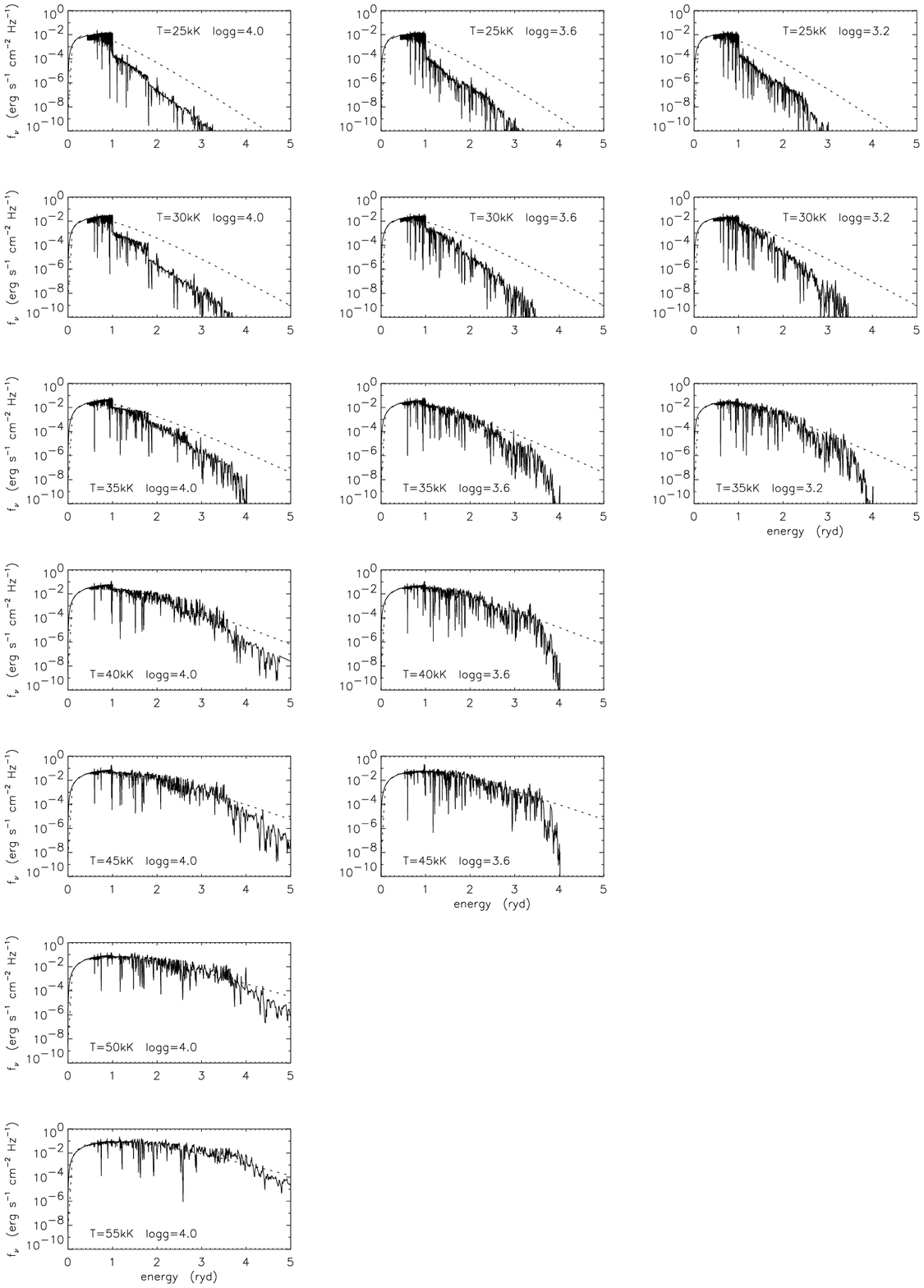}
\caption[Subsample of our model atmosphere grid, with $T_{\rm eff}$ 
ranging from 25 to 55 kK, in steps of 5kK (vertical), and log $g$ ranging from
3.2 to 4 (cm s$^{-2}$), in steps of 0.4 dex (horizontal).]
{}
\label{grid}
\end{figure}

\begin{figure}[p]
\plotone{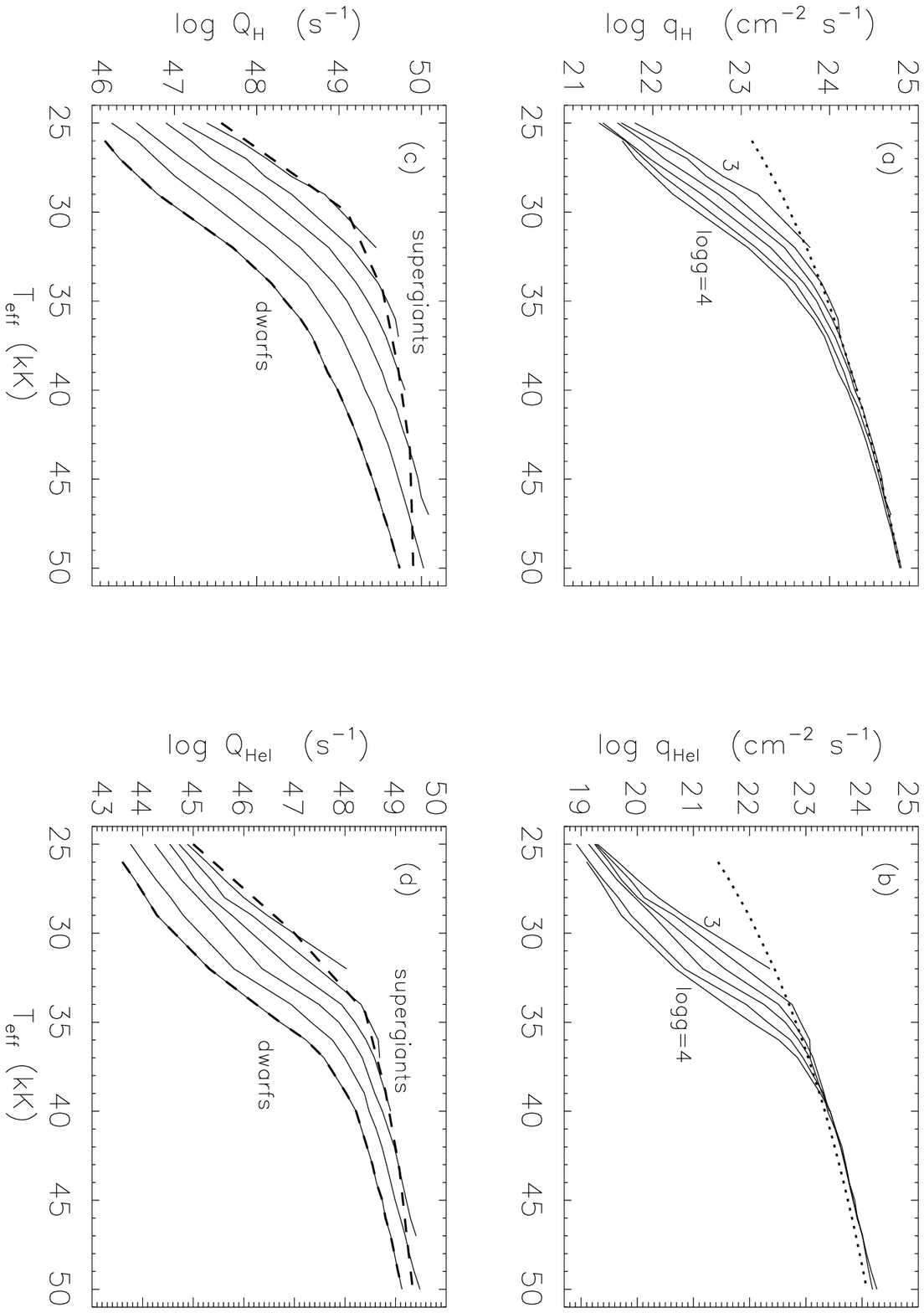}
\caption[(a) Lyman continuum surface photon flux 
$q_{\rm H}$ (cm$^{-2}$ s$^{-1}$), (b) HeI photon flux $q_{\rm He}$
(cm$^{-2}$ s$^{-1}$), (c) Lyman continuum photon 
emission rate $Q_{\rm H}$ (s$^{-1}$),
and (d) HeI photon emission rate $Q_{\rm He}$ (s$^{-1}$).
The dotted lines in panels (a) and (b) are blackbody fluxes
$\pi B_\nu(T_{\rm eff})$.
The solid curves are for stars with log $g$ ranging from 3 to, 4,
in steps of $\Delta {\rm log}g=0.2$.  The dashed curves
in panels (c) and (d) are the dwarf and supergiant sequences.]
{}
\label{qflux}
\end{figure}

\begin{figure}[p]
\plotone{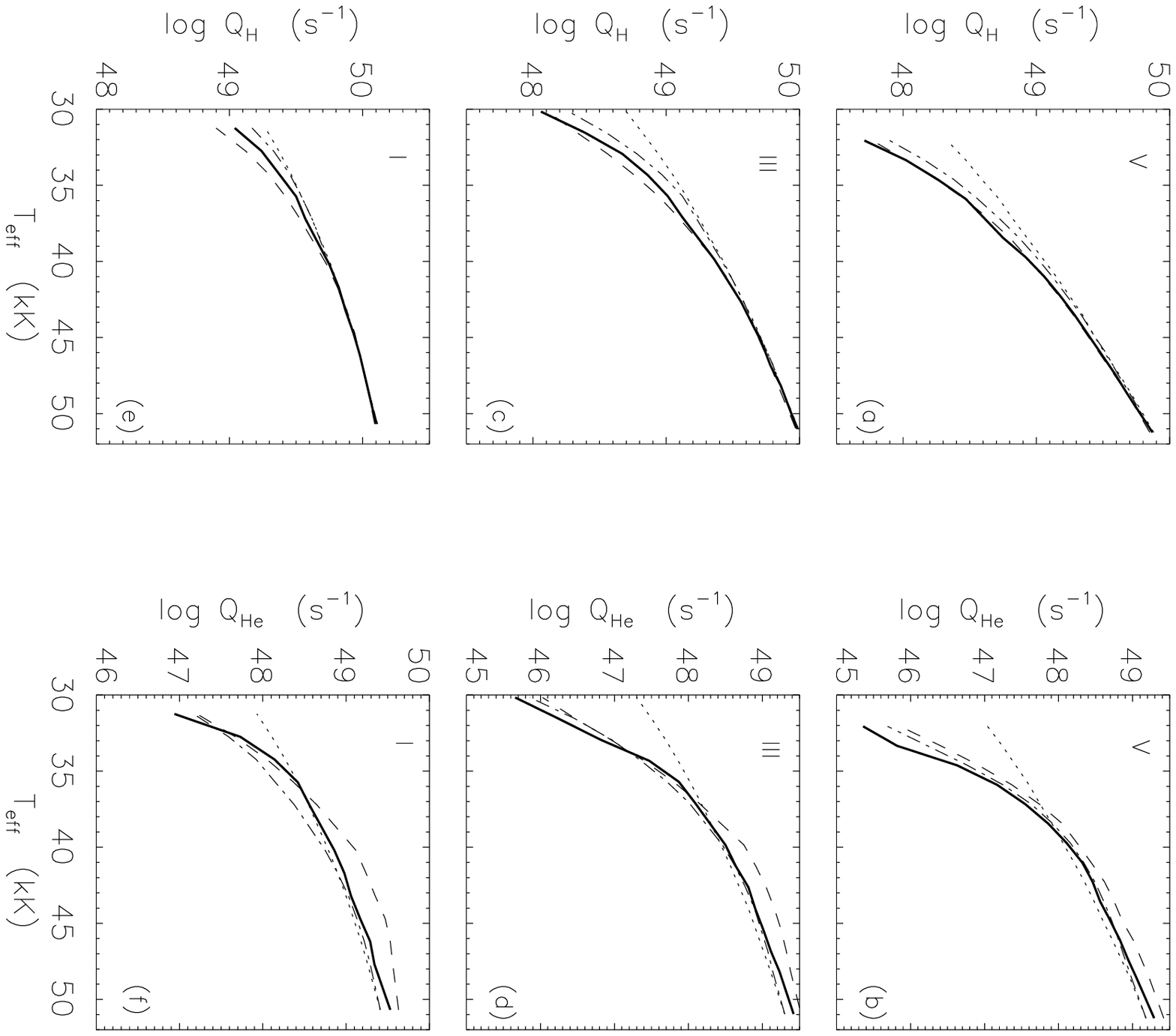}
\caption[Lyman and He {\small I} emission rate comparisons,
for dwarfs (panels a and b), giants (panels c and d),
and supergiants (panels e and f). 
Values computed in this paper are indicated by the
thick solid curves, dashed-dotted are the
Vacca et al.~(1996) rates, 
long-dashed are the CoStar
Schaerer \& de Koter (1997) rates, and dotted are the blackbody rates.]  
{}
\label{vacca}
\end{figure}

\begin{figure}[p]
\plotone{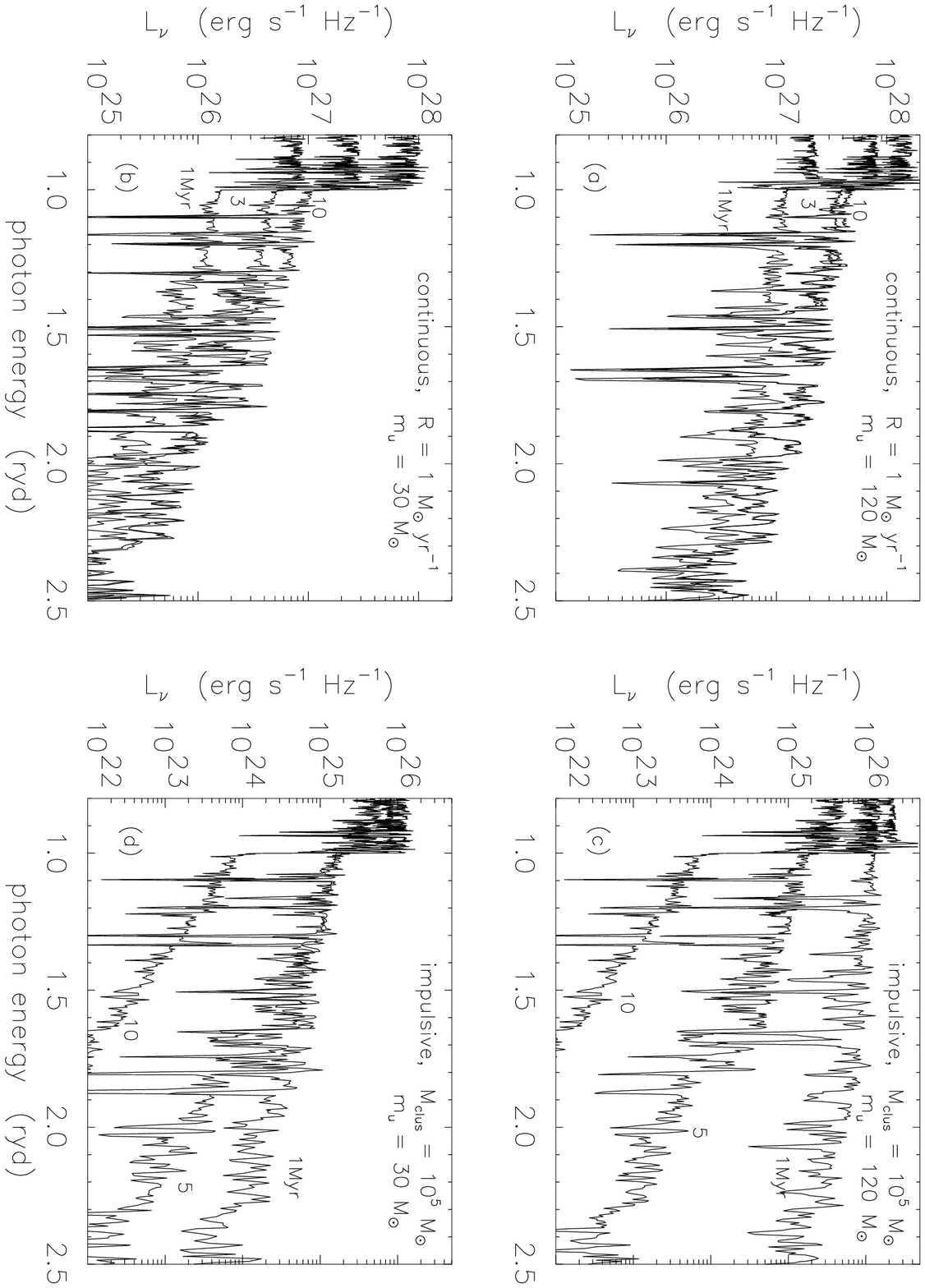}
\caption[Cluster spectral evolution. (a) Continuous star-formation
with $R=1$ M$_\odot$ yr$^{-1}$, $m_u=120$ M$_\odot$, and
cluster ages equal to 1, 3 and 10 Myr, 
(b) Continuous star-formation
with $R=1$ M$_\odot$ yr$^{-1}$, $m_u=30$ M$_\odot$, and
cluster ages equal to 1, 3 and 10 Myr, 
(c) Impulsive star-formation with cluster mass
$M_{\rm clus} =10^5$ M$_\odot$, $m_u=120$ M$_\odot$, and
cluster ages equal to 1, 5 and 10 Myr,
(d) Impulsive star-formation with cluster mass
$M_{\rm clus} =10^5$ M$_\odot$, $m_u=30$ M$_\odot$, and
cluster ages equal to 1, 5 and 10 Myr.]
{}
\label{spectra}
\end{figure}

\begin{figure}[p]
\plotone{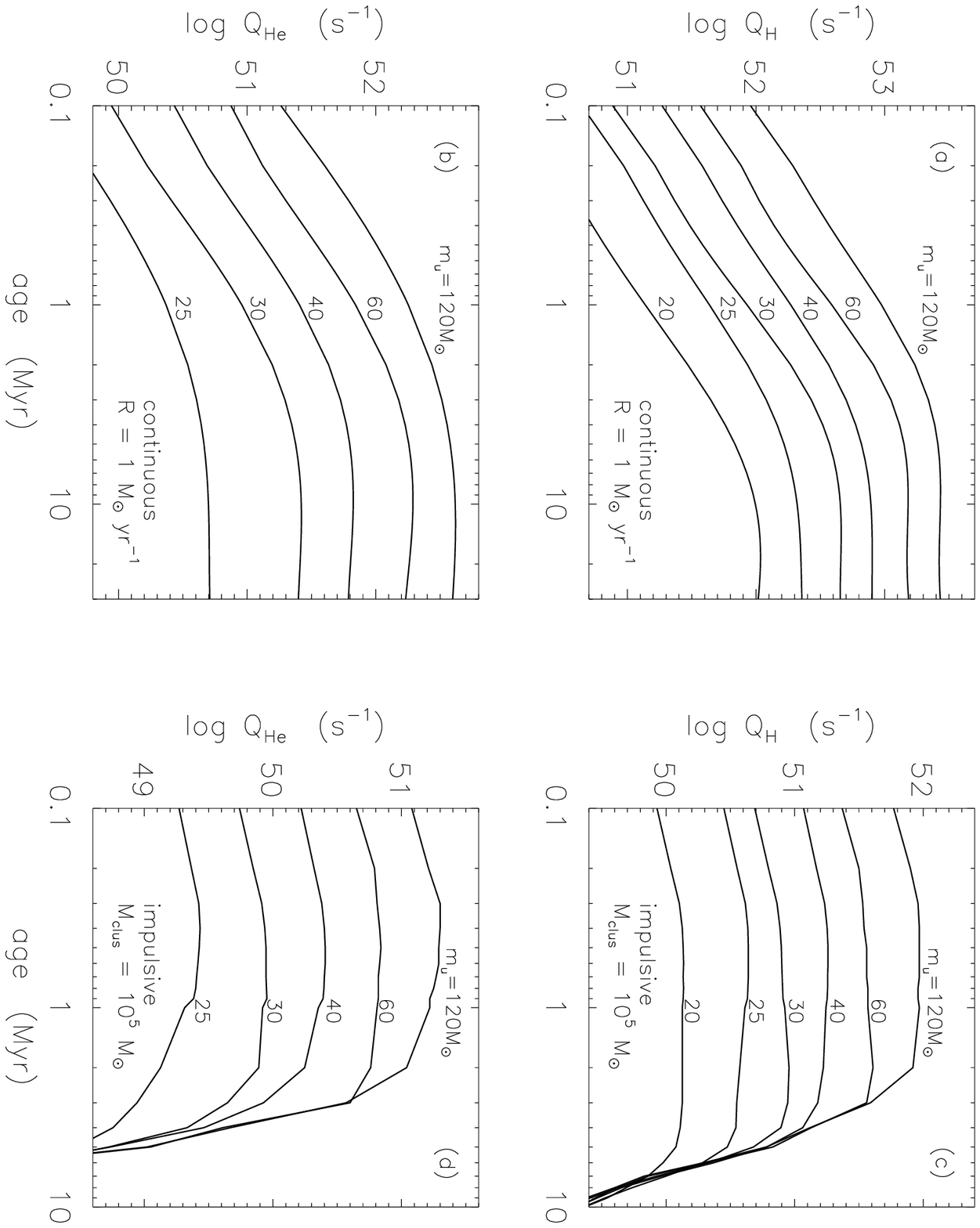}
\caption[(a) Cluster Lyman continuum photon emission rate
$Q_{\rm H}$ (s$^{-1}$), and (b) He~{\small I} emission rate $Q_{\rm He}$, 
for continuous star-formation with
$R = 1$ M$_\odot$ yr$^{-1}$. (c) Lyman emission rate
$Q_{\rm H}$, and (d) HeI emission rate $Q_{\rm He}$, for 
impulsive star-formation for a cluster mass
$M_{\rm clus} =10^5$ M$_\odot$.  In these computations a Salpeter IMF
is assumed, with an upper mass cut-off, $m_u$ ranging
from 25 to 120 $M_\odot$.] 
{}
\label{Qcluster}
\end{figure}

\begin{table}[p]
\caption{Parameters for OB stars of luminosity class V. Effective temperatures,
gravities, radii, and spectral classes, are as specified by
Vacca et al. (1996). The listed
ionizing fluxes were calculated using our model atmospheres.}
\vspace{0.5cm}
\scriptsize
\begin{tabular}{lccrcccccccc}\hline\hline
Spectral & $T_eff$ & $\log g$ & $R_{\star}$ & $\log{L}$ & $M$ & $v_\infty$ & $\dot{M}$ &
$\log{Q_H}$ & $\log{Q_{He}}$ & $\log{q_H}$ & $\log{q_{He}}$ \\
Type     & (K) & (cm $s^{-2}$) & ($R_{\odot}$) & ($L_{\odot}$)
& $(M_{\odot})$ & (km s$^{-1}$) & ($M_{\odot}$ yr$^{-1}$) & (s$^{-1}$) & (s$^{-1}$) & (cm$^{-2}$ s$^{-1}$) &
(cm$^{-2}$ $s^{-1}$) \\ \hline
O3   & 51230 & 4.149 & 13.2 & 6.04 & 87.6 & 3552 & 2.7(-6) & 49.87 & 49.29 & 24.84 & 24.26 \\
O4   & 48670 & 4.106 & 12.3 & 5.88 & 68.9 & 3266 & 1.8(-6) & 49.68 & 49.06 & 24.72 & 24.09 \\
O4.5 & 47400 & 4.093 & 11.8 & 5.81 & 62.3 & 3138 & 1.4(-6) & 49.59 & 48.94 & 24.66 & 24.01 \\
O5   & 46120 & 4.081 & 11.4 & 5.73 & 56.6 & 3026 & 1.1(-6) & 49.49 & 48.83 & 24.59 & 23.93 \\
O5.5 & 44840 & 4.060 & 11.0 & 5.65 & 50.4 & 2903 & 8.9(-7) & 49.39 & 48.70 & 24.52 & 23.83 \\
O6   & 43560 & 4.042 & 10.7 & 5.57 & 45.2 & 2784 & 7.2(-7) & 49.29 & 48.56 & 24.45 & 23.72 \\
O6.5 & 42280 & 4.030 & 10.3 & 5.49 & 41.0 & 2666 & 5.6(-7) & 49.18 & 48.46 & 24.37 & 23.65 \\
O7   & 41010 & 4.021 & 10.0 & 5.40 & 37.7 & 2543 & 4.5(-7) & 49.06 & 48.32 & 24.28 & 23.53 \\
O7.5 & 39730 & 4.006 &  9.6 & 5.32 & 34.1 & 2428 & 3.5(-7) & 48.92 & 48.11 & 24.17 & 23.36 \\
O8   & 38450 & 3.989 &  9.3 & 5.24 & 30.8 & 2313 & 2.7(-7) & 48.75 & 47.86 & 24.03 & 23.14 \\
O8.5 & 37170 & 3.974 &  9.0 & 5.15 & 28.0 & 2194 & 2.1(-7) & 48.61 & 47.55 & 23.92 & 22.85 \\
O9   & 35900 & 3.959 &  8.8 & 5.06 & 25.4 & 2083 & 1.7(-7) & 48.47 & 47.17 & 23.80 & 22.50 \\
O9.5 & 34620 & 3.947 &  8.5 & 4.97 & 23.3 & 1972 & 1.3(-7) & 48.26 & 46.63 & 23.62 & 21.99 \\
B0   & 33340 & 3.932 &  8.3 & 4.88 & 21.2 & 1853 & 1.0(-7) & 48.02 & 45.82 & 23.40 & 21.20 \\
B0.5 & 32060 & 3.914 &  8.0 & 4.79 & 19.3 & 1747 & 7.8(-8) & 47.71 & 45.36 & 23.12 & 20.77 \\
\hline
\end{tabular}
\label{LCV}
\end{table}

\begin{table}[p]
\caption{Parameters for OB stars of luminosity class III.}
\vspace{0.5cm}
\scriptsize
\begin{tabular}{lccccccccccc}\hline\hline
Spectral & $T_eff$ & $\log g$ & $R_{\star}$ & $\log{L}$ & $M$ & $v_\infty$ & $\dot{M}$ &
$\log{Q_H}$ & $\log{Q_{He}}$ & $\log{q_H}$ & $\log{q_{He}}$ \\
Type     & (K) & (cm s$^{-2}$) & ($R_{\odot}$) & ($L_{\odot}$)
& $(M_{\odot})$ & (km s$^{-1}$) & ($M_{\odot}$ yr$^{-1}$) & (s$^{-1}$) & (s$^{-1}$) & (cm$^{-2}$ s$^{-1}$) &
(cm$^{-2}$ $s^{-1}$) \\ \hline
O3   & 50960 & 4.084 & 15.3 & 6.15 & 101.4 & 3497 & 1.1(-5) & 49.98 & 49.42 & 24.83 & 24.27 \\
O4   & 48180 & 4.005 & 15.1 & 6.05 &  82.8 & 3186 & 8.4(-6) & 49.86 & 49.23 & 24.71 & 24.08 \\
O4.5 & 46800 & 3.971 & 15.0 & 5.99 &  75.8 & 3051 & 7.3(-6) & 49.78 & 49.11 & 24.64 & 23.97 \\
O5   & 45410 & 3.931 & 15.0 & 5.93 &  68.4 & 2912 & 6.3(-6) & 49.72 & 49.01 & 24.58 & 23.88 \\
O5.5 & 44020 & 3.891 & 14.9 & 5.88 &  62.0 & 2782 & 5.4(-6) & 49.64 & 48.90 & 24.51 & 23.77 \\
O6   & 42640 & 3.855 & 14.8 & 5.82 &  56.6 & 2662 & 4.6(-6) & 49.56 & 48.81 & 24.43 & 23.69 \\
O6.5 & 41250 & 3.820 & 14.8 & 5.76 &  52.0 & 2535 & 3.9(-6) & 49.46 & 48.64 & 24.33 & 23.52 \\
O7   & 39860 & 3.782 & 14.7 & 5.70 &  47.4 & 2416 & 3.3(-6) & 49.36 & 48.50 & 24.24 & 23.38 \\
O8   & 37090 & 3.700 & 14.7 & 5.57 &  39.0 & 2170 & 2.4(-6) & 49.12 & 48.09 & 24.00 & 22.98 \\
O8.5 & 35700 & 3.660 & 14.7 & 5.50 &  35.6 & 2062 & 2.0(-6) & 49.01 & 47.87 & 23.89 & 22.75 \\
O9   & 34320 & 3.621 & 14.7 & 5.43 &  32.6 & 1952 & 1.7(-6) & 48.86 & 47.47 & 23.74 & 22.35 \\
O9.5 & 32930 & 3.582 & 14.7 & 5.36 &  29.9 & 1843 & 1.4(-6) & 48.67 & 46.81 & 23.55 & 21.69 \\
B0   & 31540 & 3.542 & 14.7 & 5.29 &  27.4 & 1746 & 1.1(-6) & 48.39 & 46.24 & 23.27 & 21.12 \\
B0.5 & 30160 & 3.500 & 14.8 & 5.21 &  25.1 & 1642 & 9.2(-7) & 48.06 & 45.66 & 22.93 & 20.54 \\
\hline
\end{tabular}
\label{LCIII}
 \end{table}

\begin{table}[p]
\caption{Parameters for OB stars of luminosity class I.}
\vspace{0.5cm}
\scriptsize
\begin{tabular}{lccccccccccc}\hline\hline
Spectral & $T_eff$ & $\log g$ & $R_{\star}$ & $\log{L}$ & $M$ & $v_\infty$ & $\dot{M}$ &
$\log{Q_H}$ & $\log{Q_{He}}$ & $\log{q_H}$ & $\log{q_{He}}$ \\
Type     & (K) & (cm s$^{-2}$) & ($R_{\odot}$) & ($L_{\odot}$)
& $(M_{\odot})$ & (km s$^{-1}$) & ($M_{\odot}$ yr$^{-1}$) & (s$^{-1}$) & (s$^{-1}$) & (cm$^{-2}$ s$^{-1}$) &
(cm$^{-2}$ $s^{-1}$) \\ \hline
O3   & 50680 & 4.013 & 17.8 & 6.27 & 115.9 & 3435 & 1.6(-5) & 50.10 & 49.53 & 24.82 & 24.25 \\
O4   & 47690 & 3.928 & 18.6 & 6.21 & 104.7 & 3119 & 1.4(-5) & 50.02 & 49.34 & 24.70 & 24.02 \\
O4.5 & 46200 & 3.866 & 19.1 & 6.18 &  95.7 & 2973 & 1.3(-4) & 49.98 & 49.29 & 24.63 & 23.94 \\
O5   & 44700 & 3.800 & 19.6 & 6.14 &  86.5 & 2897 & 1.2(-5) & 49.93 & 49.17 & 24.56 & 23.80 \\
O5.5 & 43210 & 3.740 & 20.1 & 6.10 &  79.5 & 2670 & 1.1(-5) & 49.87 & 49.06 & 24.48 & 23.67 \\
O6   & 41710 & 3.690 & 20.6 & 6.07 &  74.7 & 2554 & 9.6(-6) & 49.82 & 48.98 & 24.41 & 23.57 \\
O6.5 & 40210 & 3.636 & 21.2 & 6.03 &  69.6 & 2424 & 8.7(-6) & 49.75 & 48.86 & 24.31 & 23.42 \\
O7.5 & 37220 & 3.516 & 22.4 & 5.94 &  59.2 & 2172 & 7.0(-6) & 49.57 & 48.56 & 24.08 & 23.08 \\
O8   & 35730 & 3.456 & 23.1 & 5.90 &  54.8 & 2057 & 6.3(-6) & 49.50 & 48.42 & 23.99 & 22.91 \\
O8.5 & 34230 & 3.395 & 23.8 & 5.85 &  50.6 & 1948 & 5.5(-6) & 49.37 & 48.14 & 23.84 & 22.60 \\
O9   & 32740 & 3.333 & 24.6 & 5.80 &  46.7 & 1847 & 4.8(-6) & 49.24 & 47.73 & 23.68 & 22.17 \\
O9.5 & 31240 & 3.269 & 25.4 & 5.74 &  43.1 & 1759 & 4.1(-6) & 49.04 & 46.94 & 23.44 & 21.35 \\
\hline
\end{tabular}
\label{LCI}
\end{table}

\end{document}